\documentstyle[12pt]{article}
\begin{document}
\def\ie{\hbox{\it i.e.}}        \def\etc{\hbox{\it etc.}}
\def\eg{\hbox{\it e.g.}}        \def\cf{\hbox{\it cf.}}
\def\etal{\hbox{\it et al.}}
\def\dash{\hbox{---}}
\def\cok{\mathop{\rm cok}}
\def\tr{\mathop{\rm tr}}
\def\Tr{\mathop{\rm Tr}}
\def\Im{\mathop{\rm Im}}
\def\Re{\mathop{\rm Re}}
\def\bR{\mathop{\bf R}}
\def\bC{\mathop{\bf C}}
\def\lie{\hbox{\it \$}} 
\def\partder#1#2{{\partial #1\over\partial #2}}
\def\secder#1#2#3{{\partial^2 #1\over\partial #2 \partial #3}}
\def\bra#1{\left\langle #1\right|}
\def\ket#1{\left| #1\right\rangle}
\def\VEV#1{\left\langle #1\right\rangle}
\let\vev\VEV
\def\gdot#1{\rlap{$#1$}/}
\def\abs#1{\left| #1\right|}
\def\pri#1{#1^\prime}
\def\ltap{\raisebox{-.4ex}{\rlap{$\sim$}} \raisebox{.4ex}{$<$}}
\def\gtap{\raisebox{-.4ex}{\rlap{$\sim$}} \raisebox{.4ex}{$>$}}


\def\lesssim{\mathrel{\mathpalette\vereq<}}
\def\gtrsim{\mathrel{\mathpalette\vereq>}}
\makeatletter
\def\vereq#1#2{\lower3pt\vbox{\baselineskip1.5pt \lineskip1.5pt
\ialign{$\m@th#1\hfill##\hfil$\crcr#2\crcr\sim\crcr}}}
\makeatother

\newcommand{\rem}[1]{{\bf #1}}

\renewcommand{\thefootnote}{\fnsymbol{footnote}}
\setcounter{footnote}{0}
\begin{titlepage}
\begin{center}

\hfill	  SNS-PH/98-27\\
\hfill    hep-ph/9901241\\
\hfill    \today\\

\vskip .5in

{\Large \bf
Atmospheric and Solar neutrinos in the light of the
SuperKamiokande results
\footnote
{Lecture given at the 98 Erice School. This work was supported in part by
the TMR Network under the EEC Contract
No. ERBFMRX - CT960090. }
}

\vskip .50in

Riccardo Barbieri

\vskip 0.2in
 
{\em Scuola Normale Superiore, and INFN, sezione di Pisa\\
I-56126 Pisa, Italia}
\vskip 0.05in

\end{center}

\vskip .5in

\begin{abstract}
The hierarchy $\Delta m^2_{atm} \gg \Delta m^2_\odot$ and the large
$\theta_{23}$ mixing angle, as suggested by neutrino oscillation experiments,
can be accounted for by a variety of lepton flavour models. A dichotomy
emerges: i) Models were all neutrino masses are bounded by 
$m_{atm}\equiv (\Delta m^2_{atm})^{{1 \over 2}}\approx 0.03 eV$;
ii) Models of quasi-degenerate neutrinos. It is
shown how these different patterns of neutrino masses may
arise from different lepton flavour symmetries. 
Physical implications are discussed in the various cases. 

\end{abstract}
\end{titlepage}

\renewcommand{\thepage}{\arabic{page}}
\setcounter{page}{1}
\renewcommand{\thefootnote}{\arabic{footnote}}
\setcounter{footnote}{0}

\section{Introduction}

Measurements of both solar and atmospheric neutrino fluxes provide strong
although still indirect 
evidence for neutrino oscillations \cite{neuosc}. 
The Super-Kamiokande collaboration has measured the magnitude and angular 
distribution of the $\nu_\mu$ flux originating from cosmic ray induced 
atmospheric showers \cite{SK}. Especially, but not only the 
angular distribution of the $\nu_\mu$ flux calls for an
 interpretation of the data in terms of large angle 
($\theta > 32^\circ$) neutrino oscillations, with $\nu_\mu$ disappearing to 
$\nu_\tau$ or a singlet neutrino and $\Delta m^2_{atm}$ close to
 $10^{-3} \mbox{eV}^2$. Five independent solar neutrino experiments , 
using three detection methods
\cite{sun1} 
\cite{sun10} \cite{sun11}, have measured solar neutrino fluxes which 
differ significantly from expectations. The data is consistent with 
$\nu_e$ disappearance neutrino oscillations, occuring either inside the sun,
with $\Delta m^2_\odot$ of order $10^{-5} \mbox{eV}^2$, or between the sun
and the earth, with $\Delta m^2_\odot$ of order $10^{-10} \mbox{eV}^2$.
The combination of data on atmospheric and solar neutrino fluxes 
therefore suggests a hierarchy of neutrino mass splittings:
$\Delta m^2_{atm} \gg \Delta m^2_\odot$.

Although this is the physical 
picture to which we stick in this lecture, two caveats
have to be remembered. A problem in one of the 
solar neutrino experiments or in the Standard Solar Model
could still allow comparable mass differences for 
$\Delta m^2_{atm}$ and $\Delta m^2_\odot$ \cite{BHSSW}. Furthermore, another
experimental result exists \cite{LSND}, 
interpretable as due to neutrino oscillations. 
The problem with it is that its description together with the
atmospheric and solar neutrino anomalies in terms of oscillations of the
3 standard neutrinos is impossible,  {\it even if 
$\Delta m^2_{atm} \approx \Delta m^2_\odot$ is allowed} \cite{BHSSW}.

In this lecture I consider theories with three neutrinos. Ignoring the 
small contribution to the neutrino mass matrix which gives $\Delta 
m^2_\odot$, there are three possible forms for the neutrino mass 
eigenvalues: 
\begin{eqnarray}
\mbox{``Hierarchical''} \hspace{1in} \overline{m}_\nu &=& m_{atm} 
\pmatrix{0&&\cr &0&\cr &&1} \hspace{1in} \label{eq:H} \\
\mbox{``Pseudo-Dirac''} \hspace{1in} \overline{m}_\nu &=& m_{atm} 
\pmatrix{1&&\cr &1&\cr &&\alpha} \hspace{1in} \label{eq:PD} \\
\mbox{``Degenerate''} \hspace{1in} \overline{m}_\nu &=& m_{atm} 
\pmatrix{0&&\cr &0&\cr &&1} + M \pmatrix{1&&\cr &1&\cr &&1}
\label{eq:D}
\end{eqnarray}
where $m_{atm}$ is approximately $0.03$ eV, the scale of the atmospheric 
oscillations. The real parameter $\alpha$ is either of order unity (but 
not very close to unity) or zero, while the mass scale $M$ is much larger 
than $m_{atm}$. I have chosen to order the eigenvalues so that 
$\Delta m^2_{atm} = \Delta m^2_{32}$, while $\Delta m^2_\odot = \Delta m^2_{21}$
vanishes until perturbations much less than $m_{atm}$ are added.
An important implication of the Super-Kamiokande atmospheric data is that 
the mixing $\theta_{\mu \tau}$ is large. It is remarkable that this large 
mixing occurs between states with a hierarchy of $\Delta m^2$.

What lies behind this pattern of neutrino masses and mixings? 
The conventional paradigm for models 
with flavour symmetries is the hierarchical case with hierarchically small 
mixing angles, typically given by $\theta_{ij} \approx (m_i/m_j)^{{1 
\over 2}}$. If the neutrino mass hierarchy is moderate, and if the 
charged and neutral contributions to $\theta_{atm}$ add, this kind of
 approach is probably not excluded by the data
\cite{smallangle}. It looks more interesting
to think, however, that the  neutrino masses and mixings 
do not follow this conventional pattern, since this places 
considerable constraints on model building. An 
attractive possibility is that a broken flavour symmetry leads to the 
leading order masses of (\ref{eq:H}), (\ref{eq:PD}) or (\ref{eq:D}), to 
a large $\theta_{atm}$, and to acceptable values for $\theta_\odot$ and 
$\Delta m^2_\odot$. 
To this purpose it is essential that 
the charged lepton mass matrix is discussed at the same time. 
Although it would be interesting to consider also the quark mass matrices,
this problem is only briefly mentioned here.

It turns out that it is simpler to construct flavour symmetries which 
lead to (\ref{eq:H}) or (\ref{eq:PD}) with large $\theta_{atm}$,
as illustrated in Sect 2. In both 
hierarchical and pseudo-Dirac cases, the neutrino masses have upper 
bounds of $(\Delta m^2_{atm})^{{1 \over 2}}$. In these schemes the sum of the 
neutrino masses is also bounded, $\Sigma_i m_{\nu i} \leq 0.1$ eV, implying 
that neutrino hot dark matter has too low an abundance to be relevant for 
any cosmological or astrophysical observation.
By contrast, it is more difficult to construct 
theories with flavour symmetries for 
the degenerate case \cite{degenerate}, 
were the total sum $\Sigma_i m_{\nu i} = 3M$ 
is unconstrained by any oscillation data. While non-Abelian symmetries 
can clearly obtain the degeneracy of (\ref{eq:D}) at zeroth order, the 
difficulty is in obtaining the desired lepton mass hierarchies and mixing 
angles, which requires flavour symmetry breaking vevs pointing in very 
different directions in group space. I propose a solution to this vacuum 
misalignment problem in Sect 4, which
can be used to construct a variety of models, some 
of which predict $\theta_{atm} = 45^\circ$.
Along these lines one can also construct a model with 
bimaximal mixing \cite{bimax} having $\theta_{atm} =
45^\circ$ and $\theta_{12}= 45^\circ$ \cite{deg neu}.

\section{"Hierarchical" or "Pseudo-Dirac" neutrino masses}

A large $\theta_{\mu \tau}$ mixing angle can simply be
attributed to an abelian symmetry
which does not distinguish between the muon and the tau left-handed lepton
doublets. Since this does not constrain the right handed singlets, some asymmetry
between them can be responsible 
of the $\mu-\tau$ mass difference. This is strightforward, but 
not enough, however. As emphasized
above, the $\Delta m^2_{atm} \gg \Delta m^2_\odot$ hierarchy,
if real and significant, has to be
explained as well. 

At least two textures 
for the neutrino mass matrices
have been singled out which can be responsible
for the "Pseudo-Dirac" or "Hierarchical" cases 
respectively and give at the same time a large $\theta_{atm}$ \cite{BHS} \cite{large}: 
\begin{equation}
\lambda_\nu^{(0)I} = \pmatrix{0&B&A \cr B&0&0 \cr A&0&0} \hspace{0.5in}
\lambda_\nu^{(0)II} = \pmatrix{0&0&0 \cr 0&{B^2 \over A}&B \cr 0&B&A} 
\label{eq:3nu}
\end{equation}
The important point is that both
 textures $I$ and $II$ can be obtained by the seesaw mechanism
with abelian symmetries and no tuning of parameters. 
For example, a simple model
for texture $II$ has a single heavy Majorana right-handed neutrino, $N$,
with interactions $l_{2,3}NH + MNN$, which could be guaranteed, for
example, by a $Z_2$ symmetry with $l_{2,3}, N$ odd and $l_1$ even.
A simple model for texture $I$ has two heavy right-handed neutrinos
which form the components of a Dirac state and have the interactions
$l_1 N_1 H + l_{2,3} N_2 H + M N_1 N_2$. These interactions could
result, for example, from a U(1) symmetry with $N_1, l_{2,3}$ having
charge +1, and $N_2, l_1$ having charge $-1$. In both cases, 
the missing right-handed neutrinos can be heavier, 
and/or have suitably suppressed couplings.

It makes actually sense to speak of neutrino mass textures only in 
association with corresponding charged lepton mass textures.
In turn, these textures have to be obtainable by the same symmetries 
responsible of $\lambda_\nu^{(0)I}$ or $\lambda_\nu^{(0)II}$. It is easy to
see how  $\lambda_\nu^{(0)I}$ or $\lambda_\nu^{(0)II}$ can be coupled
to 

\begin{equation}
\lambda_E^{(0)} = \pmatrix{0&0&0 \cr 0&0&B \cr 0&0&A}
\label{eq:IV}
\end{equation}
I have rotated the right-handed charged leptons so that the
entries of the first two columns vanish.
As in (\ref{eq:3nu}), I take $A$ and $B$ to be non-zero and
comparable, since they both occur consistently
with the unbroken flavour symmetries. Both in (\ref{eq:3nu}) and in 
(\ref{eq:IV}) the label $^{(0)}$ denotes the fact that
suitable perturbations have to be added to obtain fully
realistic masses and mixings.

Finally, it is immediate to extend these solutions 
to quarks, in particular to SU(5) unification, where
each field is replaced by its parent SU(5) multiplet $(l_{2,3}\rightarrow
\bar{F}_{2,3}, etc)$. Some qualitatively successful relations are
actually implied by this extension, like, e.g., $V_{cb} =O(m_s/m_b)$.

\section{Textures for quasi-degenerate neutrinos}

What are the possible textures for the degenerate case in the flavour
basis? These textures will provide the starting point for constructing
theories with flavour symmetries.
In passing from flavour basis to mass basis, the relative
transformations on $e_L$ and $\nu_L$ gives the leptonic mixing  matrix
$V$. Defining $V$ by the charged current in the mass basis,
$\overline{e} V \nu$, I choose to parameterize $V$ in the form
\begin{equation}
V = R(\theta_{23}) R(\theta_{13}) R(\theta_{12})
\label{eq:V}
\end{equation}
where $R(\theta_{ij})$ represents a rotation in the $ij$ plane by
angle $\theta_{ij}$, and diagonal phase matrices are left implicit.
The angle $\theta_{23}$ is necessarily large as it is
$\theta_{atm}$. In contrast, the Super-Kamiokande data constrains
$\theta_{13} \leq 20^\circ$ \cite{BHSSW} \cite{less20}, 
and if $\Delta m^2_{atm} > 2 \times
10^{-3} \mbox{eV}^2$, then the CHOOZ data requires $\theta_{13} 
\leq 13^\circ$ \cite{CHOOZ}. For small angle MSW oscillations in the sun,
$\theta_{12} \approx 0.05$, while other descriptions of the solar
fluxes require large values for $\theta_{12}$.

Which textures give such a $V$ together with the degenerate mass 
eigenvalues of eq. (\ref{eq:D})? 
In searching for textures, I require that in the flavour basis
any two non-zero entries are either independent or equal up to a phase, 
as could follow simply from flavour symmetries.
This allows just 
three possible textures for $m_\nu$ at leading order \cite{deg neu} 

\begin{eqnarray}
``A''  \hspace{0.25in} m_\nu & = & M \pmatrix{1 & 0 & 0 \cr 
0&1&0 \cr 0&0&1} + m_{atm} \pmatrix{0 & 0 & 0 \cr 0&0&0 \cr 0&0&1} \label{eq:A}\\
``B''  \hspace{0.25in} m_\nu & = & M \pmatrix{0 & 1 & 0 \cr 
1&0&0 \cr 0&0&1} + m_{atm} \pmatrix{0 & 0 & 0 \cr 0&0&0 \cr 0&0&1} \label{eq:B}\\
``C''  \hspace{0.25in} m_\nu & = & M \pmatrix{1 & 0 & 0 \cr 
0&0&1 \cr 0&1&0} + m_{atm} \pmatrix{0 & 0 & 0 \cr 0&1&-1 \cr 0&-1&1} \label{eq:C}
\end{eqnarray}
Alternatives for the perturbations proportional to $m_{atm}$ are
possible. Each of these textures will have to be coupled 
to corresponding suitable textures for the charged lepton
mass matrix $m_E$, defined by $\overline{e_L} m_E e_R$. For example, in cases
(A) and (B), the big $\theta_{23}$ rotation angle will have
to come from the diagonalization of $m_E$.

To what degree are the 
three textures A,B and C the same physics written in different bases, and 
to what extent can they describe different physics?
Any theory with degenerate neutrinos 
can be written in a texture A form, a texture B form or a texture C 
form, by using an appropriate choice of basis. 
However, for certain cases, the physics may be more transparent 
in one basis than in another, as illustrated later.



\section{Degenerate neutrinos from broken non-abelian symmetries}

The near degeneracy of the three neutrinos requires a non-abelian flavour 
symmetry, which I take to be $SO(3)$, with the three lepton doublets, 
$l$, transforming as a triplet. This is for simplicity -- many 
discrete groups, such as a twisted product of two $Z_2$s, would also give 
zeroth order neutrino degeneracy \cite{deg neu} \cite{SO3}. 

Following ref \cite{deg neu}, I work in a supersymmetric theory and
introduce a set of ``flavon'' chiral superfields  which 
spontaneously break SO(3). For now I just assign the desired vevs to 
these fields; later I show how 
to construct potentials which force these orientations.
Also, for simplicity I assume one set of flavon fields, 
$\chi$, couple to operators which give neutrino masses, and another set, 
$\phi$, to operators for charged lepton masses. Fields are labelled according 
to the direction of the vev, e.g. $\phi_3 = (0,0,v)$. 
For example, texture A, with 
\begin{equation}
m_E = \pmatrix{0 & 0 & 0 \cr 0&\delta_2&D_2 \cr 0&\delta_3&D_3}\equiv m_{II}, 
\hspace{0.5in}
\end{equation}
results from the superpotential 
\begin{equation}
W = (l \cdot l)hh + 
(l \cdot \chi_3)^2 hh  + (l \cdot \phi_3) \tau h  +
(l \cdot \phi_2) \tau h + (l \cdot \phi_3) \xi_\mu \mu h
 + (l \cdot \phi_2) \xi_\mu \mu h 
\label{eq:Aops}
\end{equation}
where the coefficient of each operator is understood to 
be an order unity coupling multiplied by the appropriate inverse power of 
the large flavour mass scale $M_f$.
The lepton doublet $l$ and the $\phi, \chi$ flavons are all 
$SO(3)$ triplets, while the right-handed 
charged leptons ($e, \mu, \tau$) and the Higgs doublets, $h$, are $SO(3)$ 
singlets. 
The form of eqn. (\ref{eq:Aops}) may be guaranteed by additional
Abelian flavour symmetries; in the limit where these symmetries
are exact, the only charged
lepton to acquire a mass is the $\tau$. These
symmetries are broken by vevs of flavons $\xi_{e, \mu}$, which are
$SO(3)$ and standard model singlet fields. The hierarchy of charged
fermion masses is then generated by ${ \vev{\xi_{e, \mu}} \over
M_f}$. 
The ratios $\vev{\phi_{2,3}} / M_f$ and $\vev{\chi}/M_f$ 
generate small dimensionless $SO(3)$ symmetry breaking parameters. 
The first term of (\ref{eq:Aops}) generates an $SO(3)$ invariant mass
for the neutrinos corresponding to the first term in (\ref{eq:A}). The
second term gives the second term of (\ref{eq:A}) with $m_{atm}/M =
\vev{\chi_3}^2 / M_f^2$. The remaining terms generate the charged
lepton mass matrices. Note that the charged fermion masses
vanish in the $SO(3)$ symmetric
limit --- this is the way I reconcile the near degeneracy of the
neutrino spectrum with the hierarchical charged lepton sector.

In this example we see that the origin of large $\theta_{atm}$ is due to 
the misalignment of the $\phi$ vev directions relative to that of the 
$\chi$ vev. This is generic. 
In theories with flavour symmetries, large $\theta_{atm}$ will 
always arise because of a misalignment of flavons in charged and neutral 
sectors. To obtain $\theta_{atm} = 45^\circ$, as preferred by the 
atmospheric data, requires however a very precise misalignment, which can
occur as follows. In a basis
where the $\chi$ vev is in the direction $(0,0,1)$, there should be a
single $\phi$ field coupling to $\tau$ which has a vev in the
direction $(0,1,1)$, where an independent phase for each entry is
understood. 
As we shall now discuss, in theories based on $SO(3)$,
such an alignment occurs very easily, 
and hence should be viewed as a typical expectation, and certainly 
not as a fine tuning.

Consider any 2 dimensional subspace within the $l$ triplet, and label 
the resulting 2-component vector of $SO(2)$ as $\ell = (\ell_1, \ell_2)$. 
At zeroth order in SO(2) breaking only the neutrinos of $\ell$ acquire a 
mass, and they are degenerate from $\ell \cdot \ell hh$. Introduce a flavon 
doublet $\chi = (\chi_1, \chi_2)$ which acquires a vev to break $SO(2)$.
If this field were real, then one could do an $SO(2)$ rotation to set
$\vev{\chi_2} =0$. However, in supersymmetric theories 
$\chi$ is complex and 
a general vev has the form $\vev{\chi_i} = a_i + ib_i$. 
Only one of these four real parameters can be set to zero using $SO(2)$ 
rotations. Hence the scalar potential can determine a variety of 
interesting alignments. There are two alignments which are easily 
produced and are very useful in constructing theories:
\begin{equation}
\mbox{``SO(2)'' Alignment:} \hspace{0.4in} W = X(\chi^2 - M^2); \hspace{0.4in} 
m_\chi^2 >0; \hspace{0.4in} \vev{\chi} = M(0,1).
\label{eq:so2al}
\end{equation}
The parameter $M$, which could result from the vev of some $SO(2)$ 
singlet, can be taken real and positive by a phase choice  
for the fields. The parameter $m_{\chi^2}$ is a 
soft mass squared for the doublet $\chi$.

The second example is:
\begin{equation}
\mbox{``U(1)'' Alignment:} \hspace{0.4in} W = X\varphi^2; \hspace{0.3in} 
m_\varphi^2 <0; \hspace{0.3in} \vev{\varphi} = V(1,i) \mbox{ or } V(1,-i).
\label{eq:u1al}
\end{equation}
It is now the negative soft mass squared which forces a magnitude $\sqrt{2}|V|$
for the vev.

The vev of the $SO(2)$ alignment, (\ref{eq:so2al}), picks out the original 
$SO(2)$ basis; however, the vev of the $U(1)$ alignment,
(\ref{eq:u1al}), picks out a new basis $(\varphi_+, \varphi_-)$, 
where $\varphi_\pm = (\varphi_1 \pm i 
\varphi_2)/ \sqrt{2}$. If $\vev{(\varphi_1, \varphi_2)} \propto (1,i)$, then
$\vev{(\varphi_-, \varphi_+)} \propto (1,0)$. An important feature of
the $U(1)$ basis is that the $SO(2)$ invariant $\varphi_1^2 +
\varphi_2^2$ has the form $2 \varphi_+ \varphi_-$. In the SO(3)
theory, we usually think of $(l \cdot l) hh$ as 
giving the unit matrix for neutrino masses as in texture A. However, if 
we use the $U(1)$ basis for the 12 subspace, this operator actually gives 
the leading term in texture B, whereas if we use the $U(1)$ basis in the 
23 subspace we get the leading term in texture C. 

Using this trick, it is simple to write down a variety 
of models for quasi degenerate neutrinos and charged leptons
which naturally give rise to $\theta_{23}=\theta_{atm}=45^0$ and possibly also 
to $\theta_{12}=45^o$, up to
corrections vanishing as $m_{\mu}/m_{\tau}$ or $m_{e}/m_{\mu}$
go respectively to zero \cite{deg neu}. 
In particular, quasi-degenerate neutrinos with 
$\theta_{23}=\theta_{12}=45^0$ and $\theta_{13}=0$ are known \cite{gg} to meet the
condition that makes them evading not only the upper bound on the 
absolute mass from oscillation experiments but also the constraint from
neutrinoless Double Beta decay.

\section{Conclusions}

The hierarchy $\Delta m^2_{atm} \gg \Delta m^2_\odot$ and the large
$\theta_{23}$ mixing angle, as suggested by neutrino oscillation experiments,
can be accounted for by a variety of lepton flavour models. A dichotomy
emerges. 

Models were all neutrino masses are bounded by 
$m_{atm}\equiv (\Delta m^2_{atm})^{{1 \over 2}}\approx 0.03 eV$ are easily
constructed, based on abelian flavour symmetries, even though special
attention has to be payed to the origin of 
$\Delta m^2_{atm} \gg \Delta m^2_\odot$. These models can be 
straightforwardly extended
to quarks and, in particular, to SU(5) unification.
 
On the contrary, models of quasi-degenerate neutrinos are more
likely based on non-abelian flavour symmetries. 
In the limit of exact flavour symmetry, the three
neutrinos are massive and degenerate, while the three charged leptons
are massless. Such zeroth-order masses result when the three lepton
doublets form a real irreducible representation of some non-Abelian
flavour group --- for example, a triplet of $SO(3)$. A sequential
breaking of the flavour group then produces both a hierarchy of
charged lepton masses and a hierarchy of neutrino $\Delta m^2$. 
The problem of extending these non-abelian symmetries to 
the quark sector is an open one.

An
independent indication in favour of a non-abelian flavour
symmetry may come from one or maybe two of the mixing angles
being close to $45^\circ$.
A feature of the $SO(3)$ symmetry breaking is that it 
may follow a different path
in the charged and neutral sectors, leading to a vacuum misalignment
with interesting consequences. Mixing
angles of $45^\circ$ do arise from the simplest misalignment
potentials. Such mixing can explain the atmospheric neutrino data, and
can result in rotating away the $\beta \beta_{0 \nu}$ process,
allowing significant amounts of neutrino hot dark matter.

\section*{Acknowledgements}

This work summarizes the results and the conclusions obtained by
discussing and collaborating with various people. Special tanks go
to Lawrence Hall, Alessandro Strumia and Graham Ross.

\end{document}